\documentclass[twocolumn,pra,aps,showpacs,superscriptaddress]{revtex4-1}
\usepackage{graphicx,graphics}
\usepackage{amsmath}
\usepackage{amssymb}
\usepackage{epstopdf}
\usepackage{amstext}
\usepackage{amsthm}
\usepackage{amsfonts}
\usepackage{fontenc}
\usepackage{latexsym}
\usepackage{array}
\usepackage{xfrac}
\usepackage{color}

\begin{document}

\title{Bound states of dark solitons and vortices in trapped
multidimensional Bose-Einstein condensates} 

\author{I. Morera Navarro}
\affiliation{Departament de F\'isica Qu\`antica i Astrof\'isica, 
Facultat de F\'{\i}sica, Universitat de Barcelona, E--08028
Barcelona, Spain}
\affiliation{Institut de Ci\`encies del Cosmos de la Universitat de 
Barcelona, ICCUB, E--08028 Barcelona, Spain}
\author{M. Guilleumas}
\affiliation{Departament de F\'isica Qu\`antica i Astrof\'isica, 
Facultat de F\'{\i}sica, Universitat de Barcelona, E--08028 Barcelona, Spain}
\affiliation{Institut de Ci\`encies del Cosmos de la Universitat de 
Barcelona, ICCUB, E--08028 Barcelona, Spain}
\author{R. Mayol}
\affiliation{Departament de F\'isica Qu\`antica i Astrof\'isica, Facultat 
de F\'{\i}sica, Universitat de Barcelona, E--08028 Barcelona, Spain}
\affiliation{Institut de Ci\`encies del Cosmos de la Universitat de 
Barcelona, ICCUB, E--08028 Barcelona, Spain}
\author{A. Mu\~noz Mateo}
\affiliation{Departament de F\'isica Qu\`antica i Astrof\'isica, 
Facultat de F\'{\i}sica, Universitat de Barcelona, E--08028
Barcelona, Spain}

\date{\today}

\begin{abstract}
We report on the existence and stability of multidimensional bound 
solitonic states in harmonically-trapped scalar 
Bose-Einstein condensates. Their equilibrium separation, as a 
measure of the strength of the soliton-soliton or the solitonic vortex-vortex 
interaction, is provided for varying chemical potential $\mu$. Static bound 
dark solitons are shown to be dynamically stable in elongated condensates 
within 
a range of intermediate (repulsive) interparticle-interaction strength. Beyond 
this range the snaking instability manifests during the time evolution of the 
planar solitons and produces the decay into non-stationary vortex states. A 
subsequent dynamical recurrence of solitons and vortices can be observed at low 
$\mu$. At equilibrium, the bifurcations of bound dark solitons are bound 
solitonic vortices. Among them, both two-open and two-ring vortex lines are 
demonstrated to exist with both counter- and co-rotating steady velocity 
fields. The latter flow configurations evolve, for high chemical potential, into 
a stationary 3D-chain-shaped vortex and a three vortex-antivortex-vortex ring 
sequence that arrest the otherwise increasing angular 
or linear momentum respectively. As a common feature to the bifurcated vortex 
states, their excitation spectra present unstable modes with associated 
oscillatory dynamics. In spite of this,
the family of two-open counter-rotating vortices support dynamically stable 3D 
states.
  
\end{abstract}


\maketitle

\section{Introduction}

Solitary waves are localized, nondispersive excitations that can transport 
energy and momentum in a medium. They usually show a particle-like dynamics 
that features a characteristic interaction with other solitary waves. Both the 
type of solitary waves and their interactions are determined by 
the underlying medium. Among the latter, superfluid systems made 
of ultracold quantum gases stand up as excellent playgrounds for the study of 
solitary waves. In particular, in bosonic systems with repulsive interparticle 
interactions, Bose-Einstein condensates supporting dark solitons are readily 
generated in current experiments. They have been explored by means of various 
methods, for instance by phase imprinting techniques \cite{Burger1999}, by 
engineering the hyperfine states of the same atomic species 
\cite{Becker2008}, or by producing the interference of two 
condensates 
\cite{Weller2008}. The availability of plenty of experimental data has allowed, 
on the one hand, to confront the early theoretical predictions arisen from 
one-dimensional integrable systems \cite{Tsuzuki1971,Zakharov1972}, and, on the 
other hand, to go beyond in order to explore  multidimensional solitary waves
\cite{Anderson2001,Dutton2001,Ginsberg05,Becker2013,Donadello2014,Ku2016,
Aycock2016}.

Dark solitons are unstable structures in systems with dimensions higher than 
one \cite{Derrick1964,Kuznetsov1988,Frantzeskakis2010}. But these instabilities 
can be arrested 
by the presence of an external trapping. In this case, there generally exists a 
range of parameters at low 
interparticle interaction where the dynamical 
stability of multidimensional dark solitons can be ensured 
\cite{Muryshev1999}, and therefore where they can be experimentally realized 
\cite{Weller2008}. Inside the trap, the transverse extension of a 
dark soliton (measured on the surface of minimum density) increases 
with the interparticle interaction. Beyond an interaction 
threshold the excitation of long wavelength modes on the soliton surface is 
energetically favorable, by means of which a planar soliton decays into 
localized, long-life vortex lines (with either open or ring 
configurations) \cite{Ginsberg05,Becker2013,Donadello2014,Ku2014}. 
Although this fact is a drawback for the study of multidimensional dark 
solitons, it has provided a useful mechanism for the generation of different 
solitary waves \cite{Komineas2003a,MunozMateo2014}. In this way, vortex rings 
have been observed after the decay of dark solitons in three-dimensional 
systems with isotropic trapping \cite{Anderson2001}. In elongated systems 
instead, due to the presence of perturbative noise, the unstable 
planar solitons lead eventually to a single vortex line, 
a solitonic vortex that provides a perturbative amount of angular momentum 
\cite{Brand2001}.

In order to understand the interactions between solitary waves,
the realization of states containing multiple solitary waves is required. 
To this end, states containing two solitons in 
scalar BECs have been previously considered in one dimensional 
\cite{Alfimov2007,Zezyulin2008} and quasi-one-dimensional \cite{Theocharis2010} 
settings. In these works it is assumed that the soliton decay has been 
suppressed by selecting a tight enough transverse trapping 
\cite{Kuznetsov1988,Muryshev1999,MunozMateo2014}. Under these conditions, it 
has been experimentally demonstrated \cite{Weller2008} that several solitons 
can survive moving and colliding in elongated harmonic traps during long times 
(up to seconds). 

In scalar condensates, the force experienced by a dark soliton due to the 
presence of another nearby soliton is repulsive when the inter-atomic 
interactions are local \cite{Zhao1989,Pawlowski2015,Bland2015}. As a 
consequence, in the absence of external trap, there is no static solution of 
two bound solitons. Still, analytical solutions for two moving solitons have 
been found in one-dimensional (1D) settings \cite{Akhmediev1993,Gagnon1993}. 
However, when a harmonic confinement is acting, two static dark solitons 
conveniently situated at symmetric positions around the center of the trap can 
form a bound state. This equilibrium is due to the fact that the bouyancy-like 
force experienced by the solitons in the inhomogeneous density background 
(produced by the confinement) balances the 
repulsive interaction between them \cite{Theocharis2010}. Interestingly, it has 
been recently reported that in systems with two overlapped BECs the 
inter-soliton forces can 
show either repulsive or attractive character depending on the sign of the 
inter-atomic interactions between condensates~\cite{Morera2018}.

Bound two-soliton states in 1D trapped systems show dynamical instabilities in 
a regime of inter-atomic interactions that is adjacent to the 
non-interacting limit (see Fig.~\ref{Fig:1Dfeatures})~\cite{Theocharis2010}. 
However, if the inter-atomic interaction is high enough the instability is 
suppressed and the bound solitons become stable against small perturbations. 
This situation is expected to change in multidimensional systems, since 
the increasing interaction leads to larger extensions in the spatial dimensions 
of the soliton that eventually allow for the excitation of unstable 
modes. 

Inter-vortex interactions has comparatively received much more attention, 
mainly in two-dimensional (2D) settings (see for instance 
\cite{Fetter1965,Guilleumas2001,Jezek2008,Middelkamp2011,Navarro2013,
Calderaro2017}). Contrary to the soliton case, vortex 
interactions are long range due to the velocity field that they create. In the 
referred 2D systems, and in the limit of negligible vortex cores, it has been 
shown that vortices behave as charged particles, with the circulation of their 
velocity fields playing the role of charges (see e.g. \cite{Calderaro2017}). For 
this reason, two vortices 
experience an attractive force when their circulation are opposite and a 
repulsive force when they have equal sense of circulation. 
Bound two-straight-vortex states have been observed in 2D systems with
isotropic trapping \cite{Neely2010,Freilich2010}. Both co-rotating and 
counter-rotating pairs of vortices have been realized. While the former state 
does not stay static inside a circular trap, the counter-rotating pair (or 
vortex dipole) can. The latter configuration,  
have been reported to support dynamically stable states in traps with small 
anisotropic aspect ratios \cite{Pietila2006,Stockhofe2011}, and the 
typical vortex separation has been numerically computed for the pancake-shaped 
BEC \cite{Kuopanportti2011}. A more complex dynamics is expected in 
three-dimensional (3D) systems, where vortex lines can bend in response to 
external forces 
\cite{Aftalion2001,Garcia2001,Crasovan2003,Aftalion2003,Serafini2015,
Serafini2017}.

In the present work we address the study of steady bound states 
made of dark solitons or vortices in 2D and 3D elongated condensates inside a 
harmonic trap. From now on we will use the terms bound solitonic 
states or bound solitary waves to denote such generic states. In 
particular, we look for double nonlinear excitations along the weak axis of a 
system with isotropic transverse trapping. The latter is a common arrangement 
in current experiments, where moving solitary waves can be tracked 
\cite{Ginsberg05,Becker2013, Serafini2015,Ku2016,Serafini2017}. Our starting 
point is the simplest 
excitation of the mentioned type that corresponds to a two-soliton state 
symmetrically situated around the center of the trap. As we will see, 
two vortex lines, with both straight and ring configurations and both co- and 
counter-rotating velocity fields, can be found as bifurcations for increasing 
values of the inter-atomic interaction. As special cases at high interaction, 
it is worth mentioning the 3D chain-shaped-vortex or the three vortex-ring 
configuration presented by the states in the families of two co-rotating 
straight vortices and two vortex rings, respectively. Regarding the stability 
properties, we report on small 
windows of chemical potential where stable two-dark-soliton states can be 
found. Two-straight-parallel counter-rotating 3D vortices inherit 
this stability close to the bifurcation, and, beyond a region 
of oscillatory instabilities, recover it at higher interaction. 

The rest of the paper is organized as follows. In section 
\ref{sec:System} we introduce the mean-field theoretical framework: 
the Gross-Pitaevskii equation for the condensate wave function, and the 
Bogoliubov equations for the linear excitations of the stationary bound 
solitons. Sect. \ref{sec:1D} describes 1D bound solitons 
in harmonic traps. In Sect. \ref{sec:2D} we investigate the 2D systems
that provide the main features of multidimensional bound solitonic states; we 
discuss the forces acting on the solitary waves and show characteristic time 
evolutions of two-bound-dark solitons.
In Sect. \ref{sec:3D} we study the 3D bound dark solitons 
and bound vortex lines in elongated condensates; we elaborate on the excitation 
frequencies, soliton bifurcations, and decay dynamics. To sum up, we present  
our conclusions and perspectives for future work in section 
\ref{sec:conclusions}.

\section{Mean field model}
\label{sec:System}

In the mean field regime, the dynamics of a scalar Bose-Einstein condensate 
(BEC)
can be accurately described by the Gross-Pitaevskii (GP) equation for 
the wave function $\Psi(\mathbf{r},t)$: 
\begin{align}
 i\hbar\frac{\partial}{\partial t} \Psi =\left(-\frac{\hbar^2}{2m} 
\,\nabla^2+V_{\rm trap}(\mathbf{r})+g|\Psi|^2 \right)\Psi \,,
 \label{eq:tdgpe}
\end{align}
where $g=4\pi\hbar^2 a/m$ is the interparticle interaction strength 
characterized by the scattering length $a$ and the particle mass $m$.
We consider a BEC confined by a cylindrically symmetric 
harmonic trap; $V_{\rm trap}= m (\,\omega_\perp^2 (x^2+y^2)+\omega_z^2 z^2)/2$ 
in 3D, 
and $V_{\rm trap}= m (\,\omega_\perp^2 y^2+\omega_z^2 z^2)/2$ in 2D,
with aspect ratio $\lambda=\omega_\perp/\omega_z$.
 The stationary states of Eq. (\ref{eq:tdgpe}) are $\Psi(\mathbf{r},t)=\exp(-i 
\mu 
t/\hbar)\,\psi(\mathbf{r})$, where $\mu$ is the chemical potential. The 
number of particles $N$ is fixed by normalization $\int \Psi^*(\mathbf{r},t) 
\Psi(\mathbf{r},t)\,d\mathbf{r}=N$.

The dynamical stability of the stationary states is checked by introducing 
linear modes $\{u(\mathbf{r}),v(\mathbf{r})\}$ with energy 
$\mu\pm\hbar\omega$ around the equilibrium state, i.e. 
$\Psi(\mathbf{r},t)= e^{-i\mu t/\hbar} \left[\psi+\sum_\omega(u \,e^{-i\omega 
t}+ v^*e^{i\omega t})\right]$. After substitution in the GP
Eq.~(\ref{eq:tdgpe}), and keeping terms up to first order in the modes, 
one obtains the Bogoliubov equations for the linear excitations of the 
condensate
\begin{subequations}
\begin{align}
\left( {\cal H}_L + 2g|\psi|^{2}\right) u +
g\psi^{2} v = \hbar\omega \, u \, ,
\\ 
-g(\psi^*)^2 u -\left( {\cal H}_L + 2g|\psi|^{2}\right) v
 = \hbar\omega \, v\, ,
\end{align}
\label{eq:Bog0} 
\end{subequations}
where ${\cal H}_L=-\hbar^{2}\nabla^2/2m + V_{\rm trap} -\mu$, is the linear
Hamiltonian. The existence of  frequencies $\omega$ with non-vanishing 
imaginary parts 
indicates the presence of dynamical instabilities that can produce the decay 
of the stationary state.

The stationary states have been numerically obtained by using a 
Newton continuation method based on a pseudo-spectral approach with an adapted 
basis of Fourier modes, along the axial $z$-coordinate, and Laguerre functions, 
in the transverse sections. The same pseudo-spectral approach has been used to 
solve the Bogoliubov equations from exact diagonalization. For the time 
evolution we have used a third-order Adams-Bashford scheme.
For the sake of comparison with current experiments, in what follows we
report on 2D and 3D 
condensates made of $^{87}$Rb atoms in the hyperfine state 
$|F=1,m_F=0\rangle$. The corresponding scattering length is $a=5.14\times 
10^{-9} \,$m. The numerical results are obtained for a transverse harmonic trap 
of frequency $\omega_\perp=2\pi\times200\,$~Hz and different aspect ratios.

\section{Bound solitonic states in low dimensional systems}
\label{sec:1D}
\begin{figure}[tb]
\centering
\includegraphics[width=\linewidth]{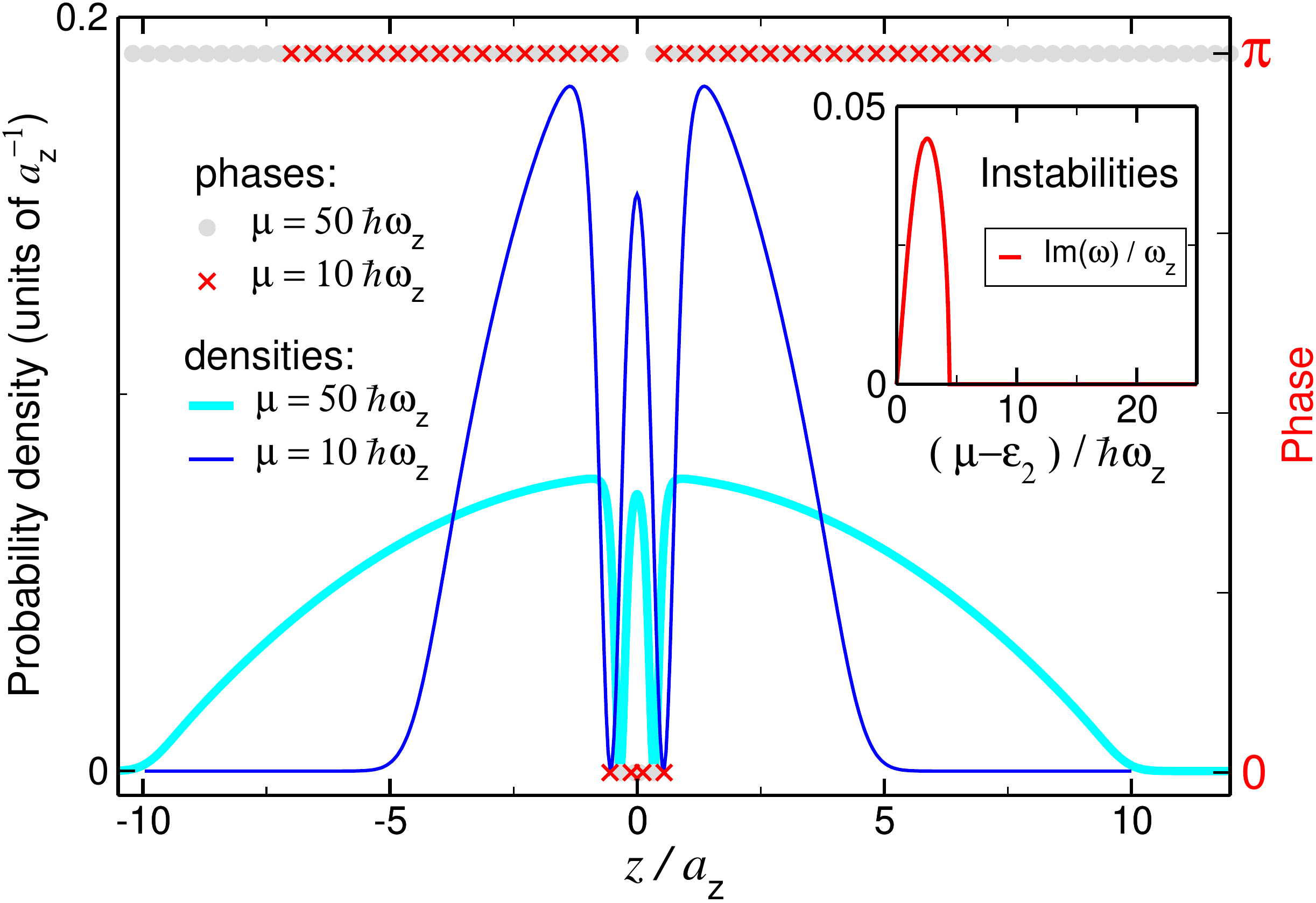}\\
\vspace{0.1cm}
\includegraphics[width=\linewidth]{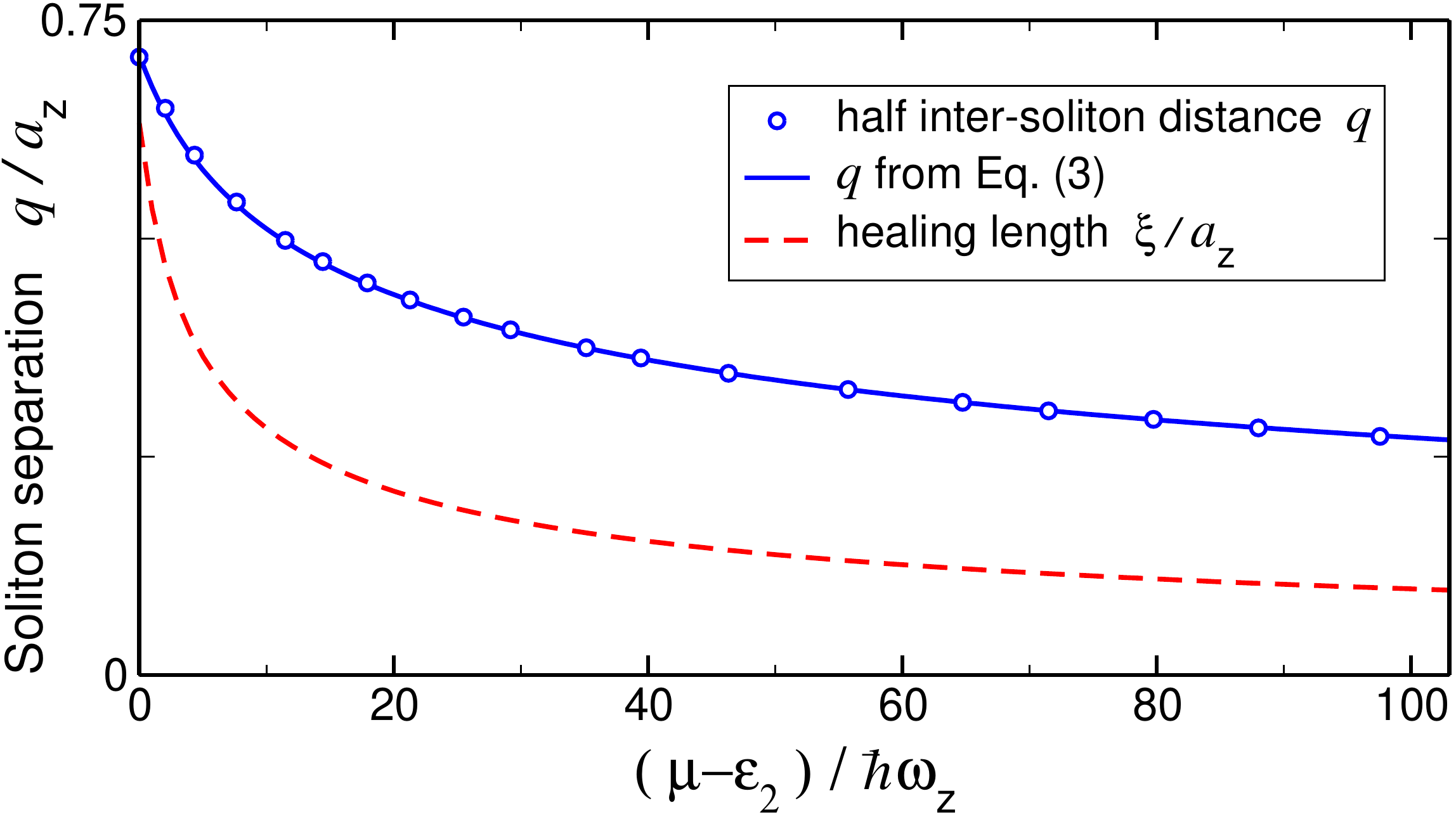}
\caption{Features of stationary two-bound-soliton states in harmonically 
trapped 1D BECs. Top panel: Probability density $|\psi|^2/N$ (lines) and phase 
$\arg(\psi)$ (symbols) for two values of the chemical potential $\mu$.
The inset shows the imaginary frequencies of unstable 
excitation modes in the two-bound-soliton family. Bottom panel: (half) soliton 
separation $q$ as a function of $\mu$, measured relative to the energy of the 
second excited linear eigenstate $\varepsilon_2=2.5\,\hbar\omega_z$. 
The analytical expression (\ref{eq:1Dseparation}) (solid line) follows 
in very good approximation the numerical data from the solution of the GP 
equation (open symbols). The healing length $\xi$ (dashed line) is shown for 
comparison.}
\label{Fig:1Dfeatures}
\end{figure}

It is instructive to start with the analysis of pure 1D systems.
In the non-interacting regime, the second excited axial eigenmode
of the 1D harmonic oscillator is the second Hermite polynomial
$H_2=[2(z/a_z)^2-1]\exp(-z^2/2a_z^2)/2\sqrt{a_z\sqrt{\pi}}$, which has
two axial nodes symmetrically situated around the trap center at distance $q 
\equiv  z_{node}= a_z/\sqrt{2}$, where 
$a_z=\sqrt{\hbar/m\omega_z}$. The nonlinear continuation (for $g>0$) of this 
state keeps the topology (the same number of nodes) and builds the family of 
two-bound-dark solitons in the harmonic trap.  The equilibrium distance 
between nodes is maximum at $g=0$. Two representative examples are depicted in 
the top panel of Fig. \ref{Fig:1Dfeatures}, 
containing data, density (lines) and phase (symbols), obtained from the 
numerical solution of the time-independent GP equation for chemical potentials 
$\mu=$10 and 50 $\hbar\omega_z$.

In  a system with repulsive inter-atomic interactions, dark solitons experience 
also repulsive forces between themselves. In the absence of 
external trap, it has been shown that this force decays exponentially 
with the soliton separation $2q$ as  
$F_{int}\propto \exp{(-4 q/\xi)}$ \cite{Zhao1989,Theocharis2010}, where 
$\xi=\hbar/\sqrt{m g\,n }$ is the healing length, and prevents the existence of 
bound states. In harmonically-trapped 1D systems, however, two equal dark 
solitons symmetrically situated at a distance $z=q$ from the trap center find 
an equilibrium configuration. From a particle-like approach for the soliton 
dynamics \cite{Scott2011,MunozMateo2015}, the soliton separation is 
determined at equilibrium by the balance between the inter-soliton force 
$F_{int}(z)=-16 N_s \hbar^2 \exp{(-4 z/\xi)}/m \xi^3$ and
the buoyancy-like force $F_b(z)=N_s m  \omega_z^2 z$ due to 
the inhomogeneous density background induced by the trap, where 
 $\xi$ is evaluated at maximum density, and $N_s$ ($<$0) 
is the number of particles depleted by the soliton. 
The equilibrium leads to a transcendental equation for $\tilde q=q/\xi$
\begin{align}
 \tilde q\,e^{4\tilde q}=\left(\frac{4\mu}{\hbar\omega_z}\right)^2,
 \label{eq:1Dseparation}
\end{align}
which, as can be seen in the bottom panel of Fig. \ref{Fig:1Dfeatures}, 
provides a very good estimate of the soliton separation as obtained from the 
numerical solution of the 1D GP equation. As the chemical potential 
(hence the inter-atomic interaction) increases, the healing length decreases 
and so does the soliton separation in absolute (harmonic oscillator, $a_z$) 
units. 
In healing length units, however, the inter-soliton distance $q$ increases with 
the chemical potential, which reflects a lower underlying force of 
soliton-soliton interaction.

\subsection{2D Bound solitonic states}
\label{sec:2D}
\begin{figure}[tb]
	\centering
	\includegraphics[width=\linewidth]{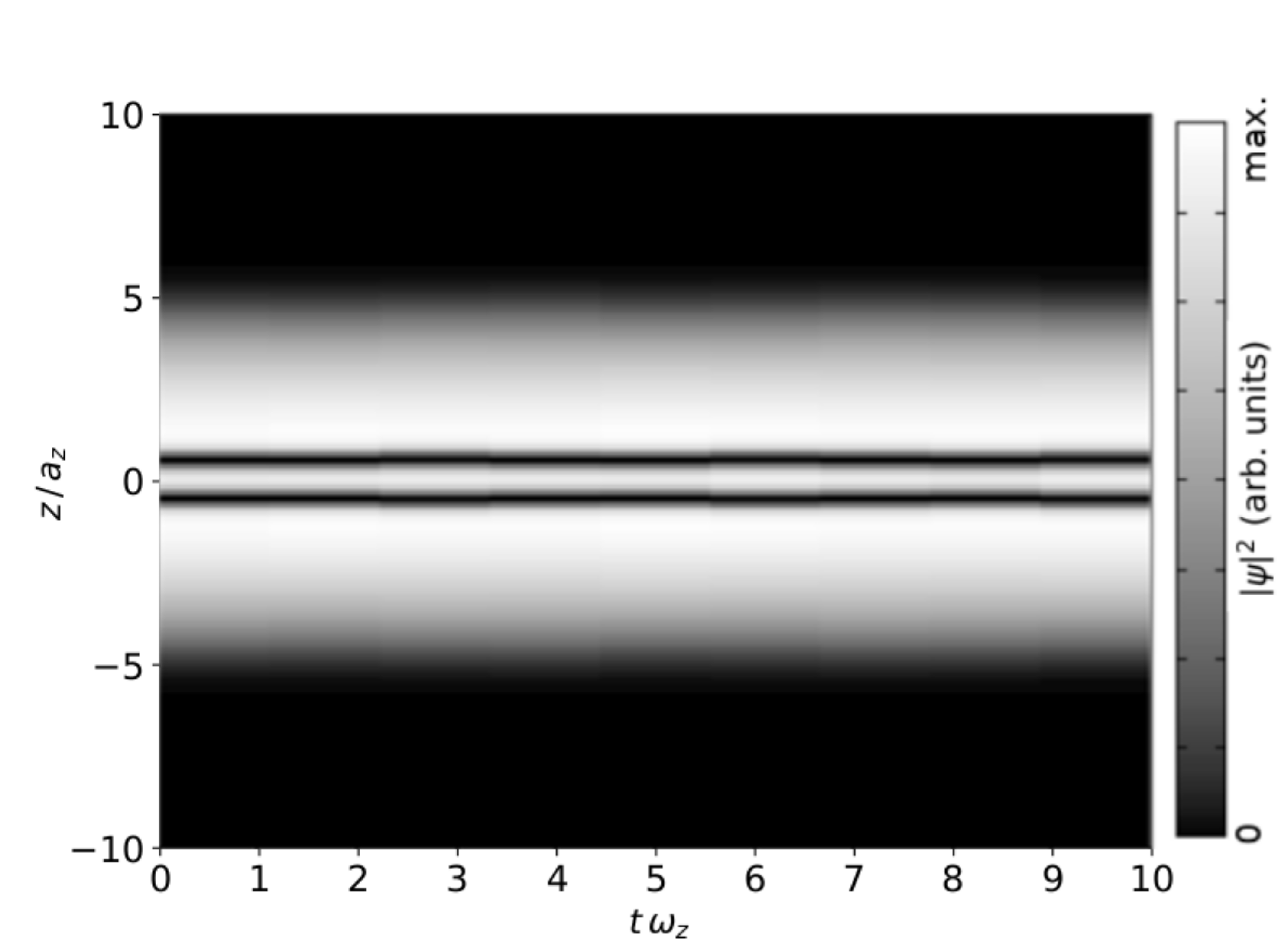}
	\includegraphics[width=\linewidth]{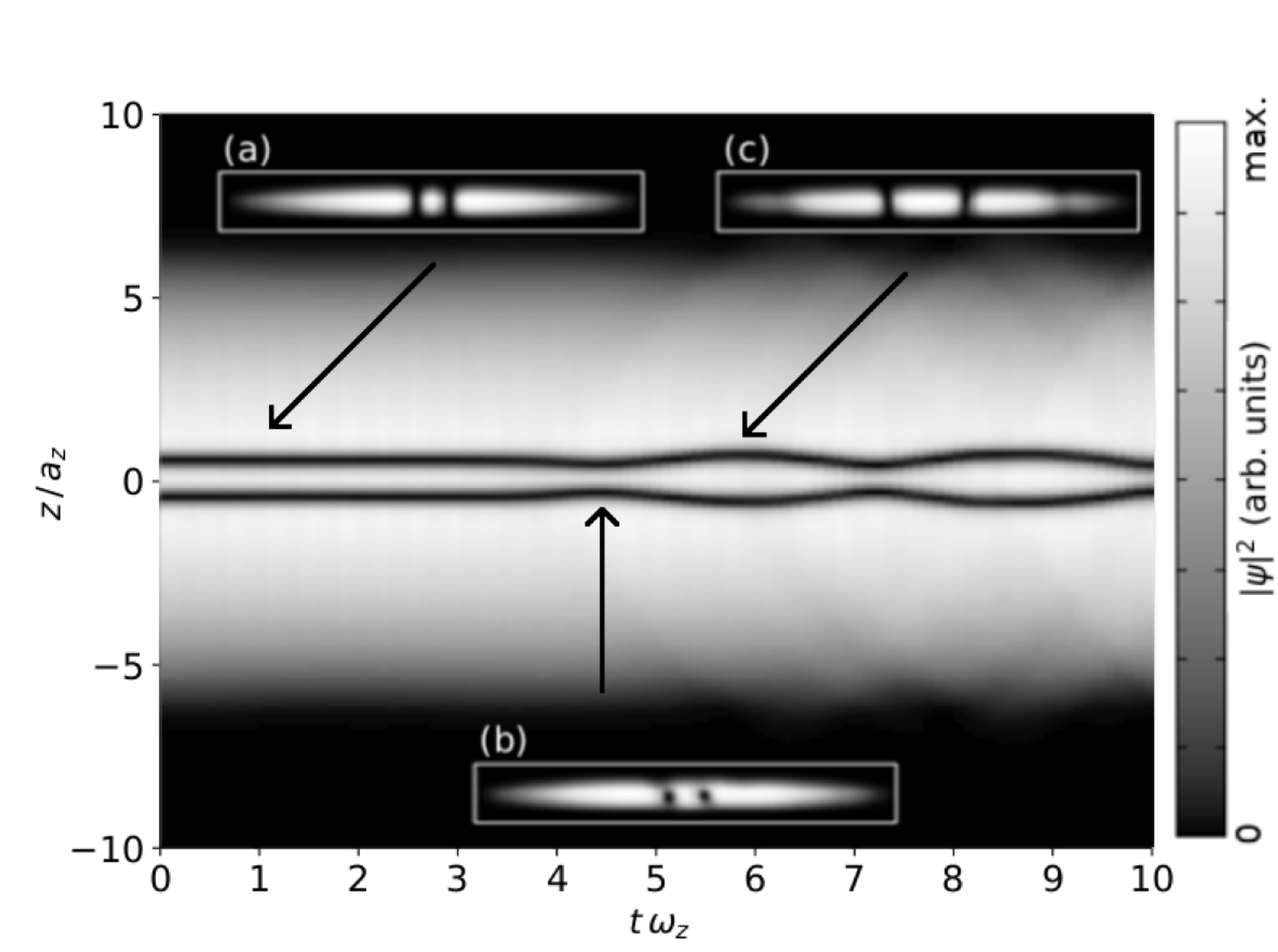}
	\caption{Time evolution of bound dark solitons across the $z$ direction 
in 2D trapped systems with aspect ratio $\lambda =10$. Sections of the density 
profile at $y=0$ are represented versus time. Top panel: robust state against 
perturbations in a BEC with $\mu = 2.2 \,\hbar\omega_\perp$. Bottom panel: 
soliton-vortex decay in a BEC with $\mu = 3 \,\hbar\omega_\perp$. The three 
insets represent snapshots of the density profile on the $z$-$y$ plane
at three different stages of the evolution: (a) initial two-soliton state, (b) 
decay into two solitonic vortices and, (c) recurrence of dark solitons.}
	\label{Fig:2Devol}
\end{figure}

The family of 1D two-bound-dark solitons contains dynamically unstable states 
for low values of the chemical potential (see the inset in the top panel of 
Fig. \ref{Fig:1Dfeatures}), where out of phase excitation modes 
break the static configuration \cite{Theocharis2010}. In multidimensional 
systems new instabilities appear. The equilibrium states are unstable against 
the bending of the dark soliton stripe (in 2D) or the dark soliton plane (in 
3D) when their extensions are long enough to support long wavelength (above a 
few healing lengths) transverse modes \cite{Muryshev1999}. However dynamically 
stable bound solitons can still be found in multidimensional settings, just in 
between the end of the out-of-phase 1D instability and the beginning of the 
snaking instability.
As an example, the top panel of Fig.~\ref{Fig:2Devol} shows a robust 2D
state with $\mu = 2.2 \, \hbar\omega_\perp$ in a trap with aspect ratio $\lambda 
= 10$. After seeding a perturbative small amount of white noise on the 
stationary state, the equilibrium configuration is preserved for a long time 
evolution.

This situation changes for higher chemical potentials (or equivalently 
for higher inter-atomic interactions), where the snaking instability can 
operate. To illustrate this process, we have prepared an initial 2D state with 
$\mu = 3 \,\hbar\omega_\perp$ having two dark solitons at their equilibrium 
distance. As can be seen in the bottom panel of Fig.~\ref{Fig:2Devol}, 
(again after adding a small white noise on the initial state) the real time 
evolution shows 
the soliton decay into counter-rotating vortices (inset (b)) that oscillate 
around their center of mass. During the motion along the weak trap  $z$-axis, 
the vortices transit through regions of lower local chemical potential where 
there is not enough energy to support vortex structures. As a result, close to 
the turning points, the vortices smoothly transform back into dark solitons 
(inset (c)). The opposite behavior, soliton to vortex conversion, occurs in the 
proximity to the trap center. This phenomenon of dynamic 
inter-conversion 
between stationary states \cite{Molina2001,Pietila2006} is 
another instance of a nonlinear 
recurrence in the GP equation that reflects the dynamical instability 
of the involved stationary states \cite{Kuznetsov2017}. However, at higher 
chemical potential, away of this oscillatory instability, robust states made of 
counter-rotating vortices can be found again \cite{Pietila2006}.

\begin{figure}[tb]
\centering
\includegraphics[width=0.46\linewidth]{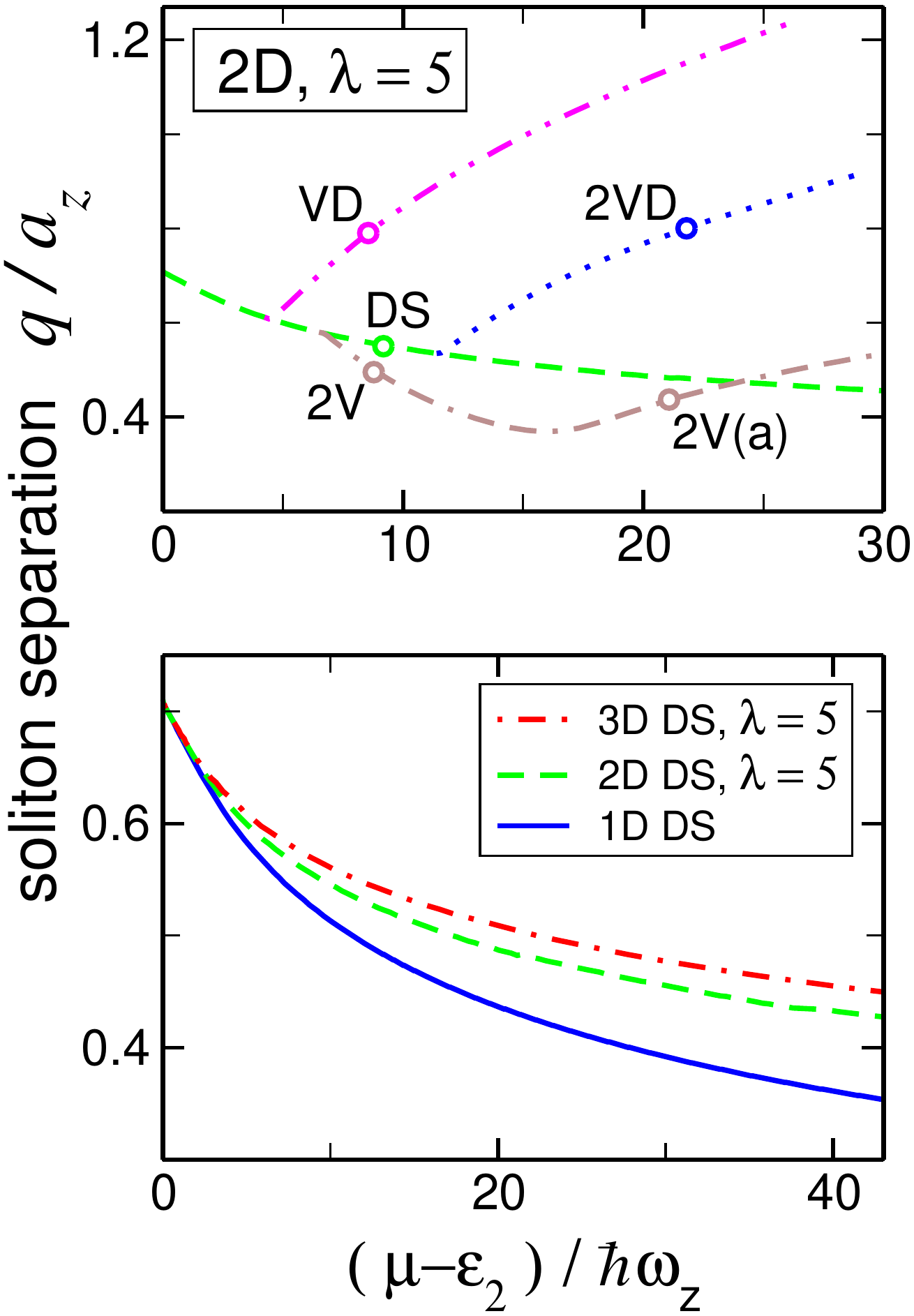}
\includegraphics[width=0.50\linewidth]{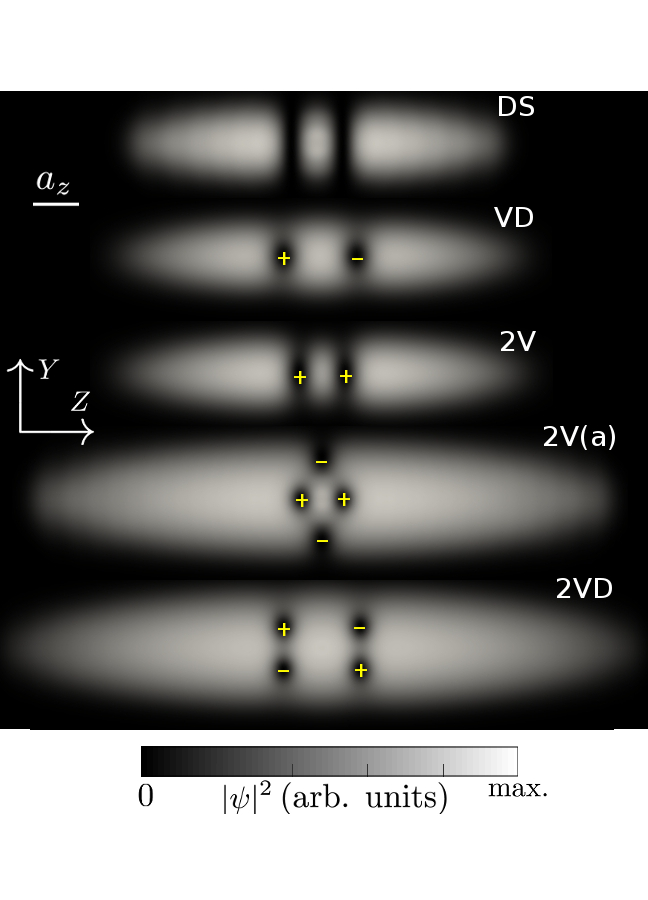}
\caption{Bound solitary waves in elongated 2D BECs. Left panels: 
half axial separation between solitary waves of different type (top) and 
between dark solitons in different dimensional settings (bottom) as a 
function of the chemical potential (measured relative to the 
energy of the second excited non-interacting eigenstate along the weak-trap 
 $z$-axis $\varepsilon_2=d\times2.5\,\hbar\omega_z$, where $d$ is the number of 
dimensions). Right panel: density in the $z$-$y$ plane of the 2D bound solitary 
waves (see text) marked by open symbols in the top left panel. 
The $\pm$ signs indicate the vortex polarity. }
\label{Fig:2Dsolitons}
\end{figure}
In elongated condensates, it has been shown that static states made 
of solitonic vortex lines bifurcate from a single dark soliton excitation 
\cite{MunozMateo2014}. A priori, one could expect a similar scenario to happen 
by starting with two dark solitons, but this is only the case if the dark 
solitons are far away from each other. Otherwise, the interaction between the 
emerging solitary waves play 
a decisive role in selecting the possible stationary states in the trap. Due to 
the different nature of vortex-vortex interaction, the distance 
of equilibrium between two bound vortices is generally different from the 
distance between two bound dark solitons (which is at the origin of the 
oscillations observed in the lower panel of Fig. \ref{Fig:2Devol}). In addition, 
the dimensionality of the system itself, modulating the underlying density 
profiles, makes also this distance to change even for the same type of solitary 
waves. 

The effect of both factors, type of solitary wave and dimensionality, on the 
solitary-wave separation in 2D systems is shown across the panels of Fig. 
\ref{Fig:2Dsolitons}. First of all, three new static bound-vortex states are 
shown to bifurcate for increasing chemical potential. As depicted in the 
right panel, they correspond to two counter-rotating vortices (or vortex 
dipole, VD), two co-rotating vortices (2V), and a double vortex dipole (2VD). 
The axial separations between the corresponding topological defects (plotted in 
the top left panel of Fig. \ref{Fig:2Dsolitons}) measure the 
associated force of interaction.

Similarly to 1D solitons, co-rotating vortices present repulsive interactions 
in homogeneous-density backgrounds. The difference resides in the range of the 
interaction, which is long range  for vortices (Coulomb type in 2D 
\cite{Calderaro2017}). 
However, in elongated settings the solitonic character of the vortices makes 
their features localized \cite{Brand2001} and the interaction turns essentially 
local \cite{Serafini2015}.  
As shown in Fig. \ref{Fig:2Dsolitons}, the separation distance 
decreases with the increasing chemical potential (initially) faster along the 
co-rotating (2V) family than along the dark soliton family. Therefore 
the vortex-vortex interaction, for given chemical potential in this regime, is 
weaker than the 
soliton-soliton interaction. For higher chemical potential, a striking 
difference arises in the former family, 
which reaches a minimum vortex separation. This is due to the entry of 
a pair of new co-rotating vortices with opposite circulation to the original 
vortices. As can be seen in the profile 2V(a) of the right panel of Fig. 
\ref{Fig:2Dsolitons}, the entry is symmetric on the perpendicular bisector of 
the line joining the 
original vortices. As a consequence the angular momentum, which kept increasing 
with decreasing vortex separation, reaches a maximum value near the minimum 
separation. Note that the angular momentum is not a conserved quantity along 
each family of stationary solitonic states. Only at the bifurcation point, 
where two solitonic families merge, all the dynamical properties are common. 
Beyond this point, the states belonging to a particular family change generally 
their properties for varying chemical potential (which is the nonlinear 
continuation parameter).

The scenario is quite different for two bound counter-rotating vortices (or 
vortex dipole, VD). In this case, the vortex-antivortex pair produces zero 
angular momentum, and presents attractive interactions on homogenous density 
backgrounds. Again, the finite size of both the system (due to the trap) and 
the vortex cores induces an effective repulsive interaction that allows 
for a bound configuration. Contrary to the states considered before, the 
vortex-antivortex separation increases with the chemical potential (see Fig. 
\ref{Fig:2Dsolitons}), denoting higher interaction forces. 
Interestingly, this behavior contrasts with the roughly constant 
(but slightly decreasing) vortex separation found in pancake geometries 
\cite{Kuopanportti2011}.

The effect of dimensionality on the bound state configuration (hence on the 
inter solitary wave interaction) is illustrated for bound dark solitons (DS) in 
the bottom left panel of Fig. \ref{Fig:2Dsolitons}. It can be seen that 
for a given chemical potential the inter-soliton distance increases with the 
number of dimensions. The cause is the inhomogeneous density profile along 
the transverse directions, which has larger healing lengths at lower local 
chemical potential and contributes with higher buoyancy forces to the overall 
interaction. More importantly, as we show below, dimensionality has striking 
consequences in 3D configurations due to the bending of vortex lines.

\section{3D bound solitonic states}
\label{sec:3D}

\begin{figure}[tb]
\centering
\includegraphics[width=0.9\linewidth]{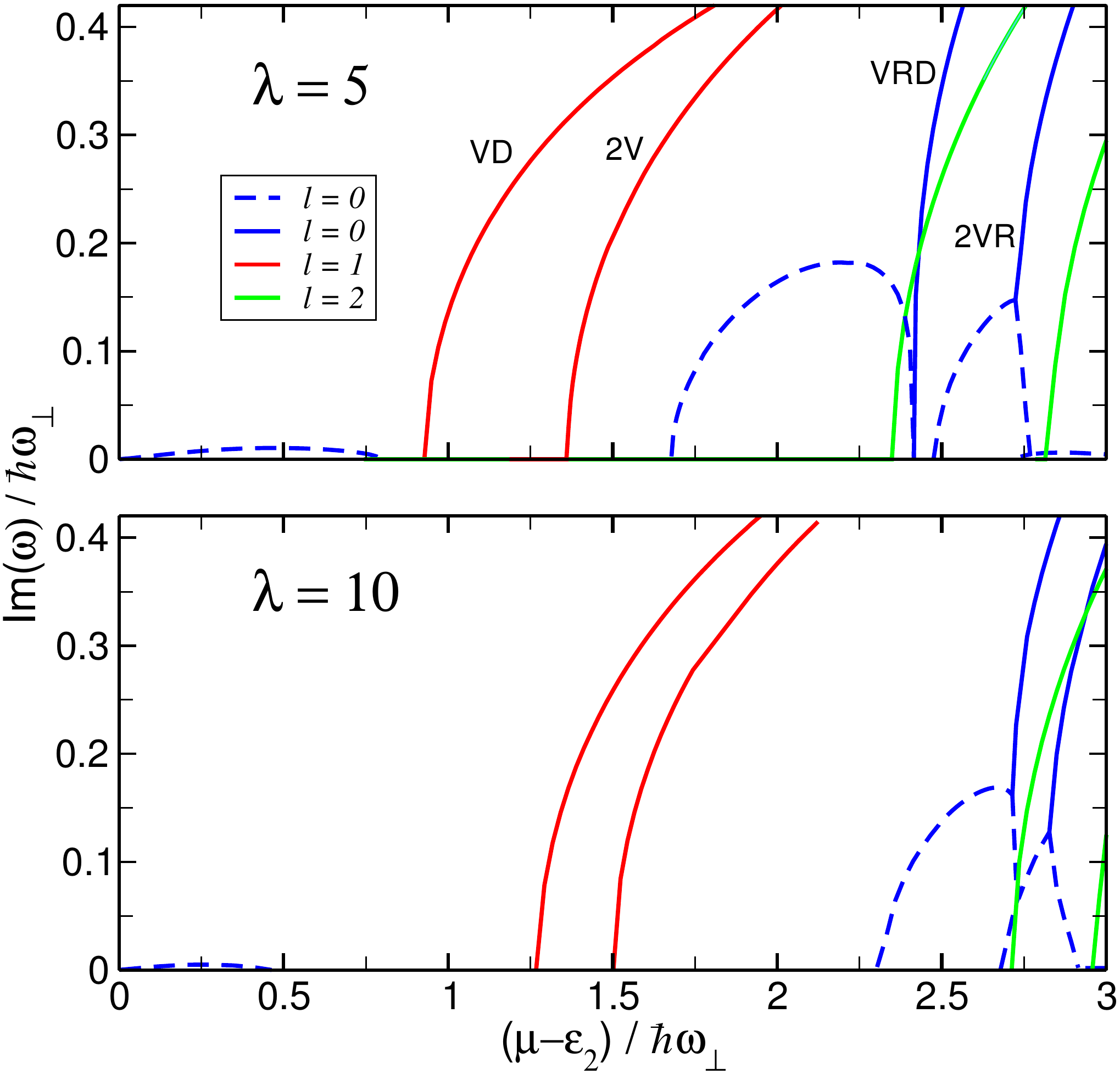}
\caption{Unstable frequencies of 3D two-bound-dark-soliton states classified 
by the azimuthal polarity $l$ (see text) of the corresponding excitation mode. 
Solid lines correspond to pure imaginary $\omega$ frequencies, whereas dashed 
lines correspond to complex frequencies. The instabilities appear by pairs for 
given $l$ according to the axial ($z$) parity of the modes. For larger aspect 
ratio $\lambda$ the stability window increases. The labels at the top 
panel indicate the nonlinear bifurcations described in the text.}
\label{Fig:3Dimag}
\end{figure}
It is convenient to start the search for families of bound solitary waves by 
considering the static state made of two bound dark solitons. In the 
non-interacting 3D case such state is given by 
$\Psi_2(x,y,z,t)=\exp(-\varepsilon_2 
t/\hbar) \exp[-(x^2+y^2)/2a_\perp^2]/\sqrt{2\pi a_\perp^2}H_2(z/a_z)$, with  
energy $\varepsilon_2=\hbar(\omega_\perp +\omega_z/2)$, where 
the axial part corresponds to the second Hermite function $H_2(z/a_z)$ of 
the 1D system. The continuation of this solution in the nonlinear regime gives 
the family of 3D bound solitons. Figure \ref{Fig:3Dimag} shows the 
frequencies of unstable modes in their excitation spectrum, obtained from the 
numerical solution of the Bogoliubov equations (\ref{eq:Bog0}). The modes are 
grouped by the azimuthal polarity $l$, a positive integer that indicates 
the number of nodal diameters on the transverse section. As can be seen,
the 3D bound solitons inherit the 1D ($l=0$) out-of-phase instability 
\cite{Theocharis2010}, but it is strongly suppressed at high aspect ratios 
$\lambda\gg1$.
More relevant for the bifurcation of new solitonic states, there are
transverse excitations leading to the so-called snaking instability that 
bends the soliton plane \cite{Kuznetsov1988,Muryshev1999}. This instability 
appears beyond a chemical potential threshold (or equivalently an 
inter-atomic interaction threshold) that marks the excitation of the 
lowest energy transverse mode at the soliton planes. Such mode has $l=1$ and 
presents a single transverse nodal line at each soliton plane
\cite{MunozMateo2014}. It can 
be viewed as the linear superposition (with equal weight) of a transverse 
vortex and an antivortex with angular momentum $L_z=\hbar$ and $L_z=-\hbar$ 
respectively. The subsequent 
growth of this unstable mode produces a solitonic vortex \cite{Brand2001} on 
the corresponding soliton planes. For this reason, the 
threshold for the excitation of the lowest transverse mode coincides with the 
bifurcation of a nonlinear wave made of solitonic vortices. The 
threshold increases with the condensate aspect ratio, tending to the limit case 
of no axial trapping, where it takes the value 
$\mu\approx2.65\hbar\omega_\perp$ \cite{Komineas2003,MunozMateo2014}.

In between the two mentioned instabilities, for an intermediate range of the 
chemical potential, our results demonstrate that there exist 
dynamically stable states made of two bound dark solitons. The 
stability window enlarges with the trap aspect ratio due to the shift
of the snaking instability threshold towards higher $\mu$ and the 
suppression of the 1D-out-of-phase instability. Therefore, the states made 
of two static bound dark solitons are susceptible of observation in current 
experiments. 
\begin{figure}[tb]
\centering
\includegraphics[width=\linewidth]{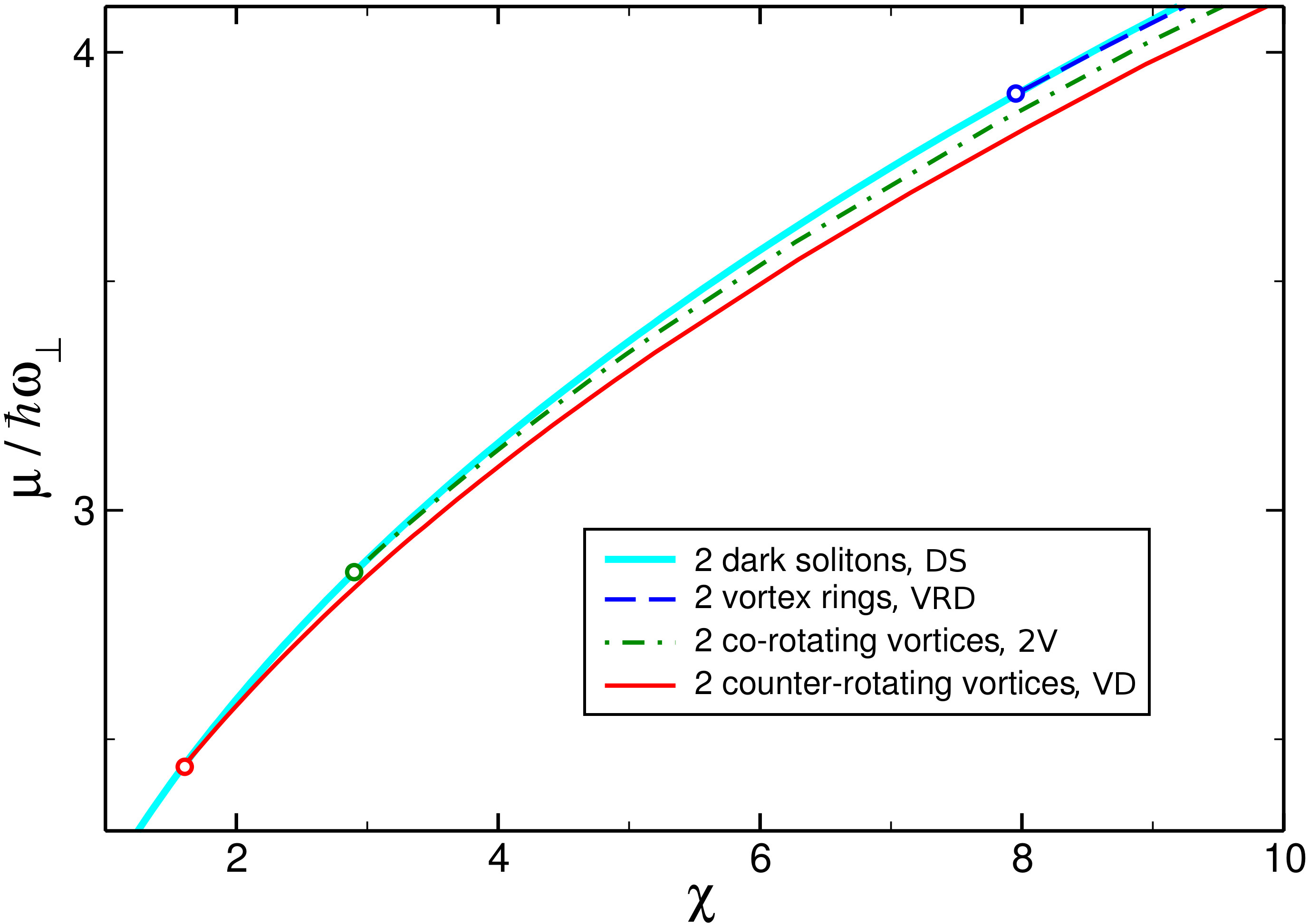}
\caption{Bifurcation of first 3D bound solitary waves in an elongated harmonic 
trap with aspect ratio $\lambda=5$. The chemical potential 
$\mu$ (in energy units of the transverse trap $\hbar\omega_\perp$) is 
represented against the interaction parameter $\chi$. The 
bifurcation points (open symbols) coincide with the corresponding emergence of 
a linear excitation mode (see Fig. \ref{Fig:3Dimag}), with equal azimuthal 
polarity $l$ and pure imaginary frequency $\omega$, on the soliton planes.}
\label{Fig:bifurcation}
\end{figure}

\begin{figure}[tb]
\centering
\includegraphics[width=\linewidth]{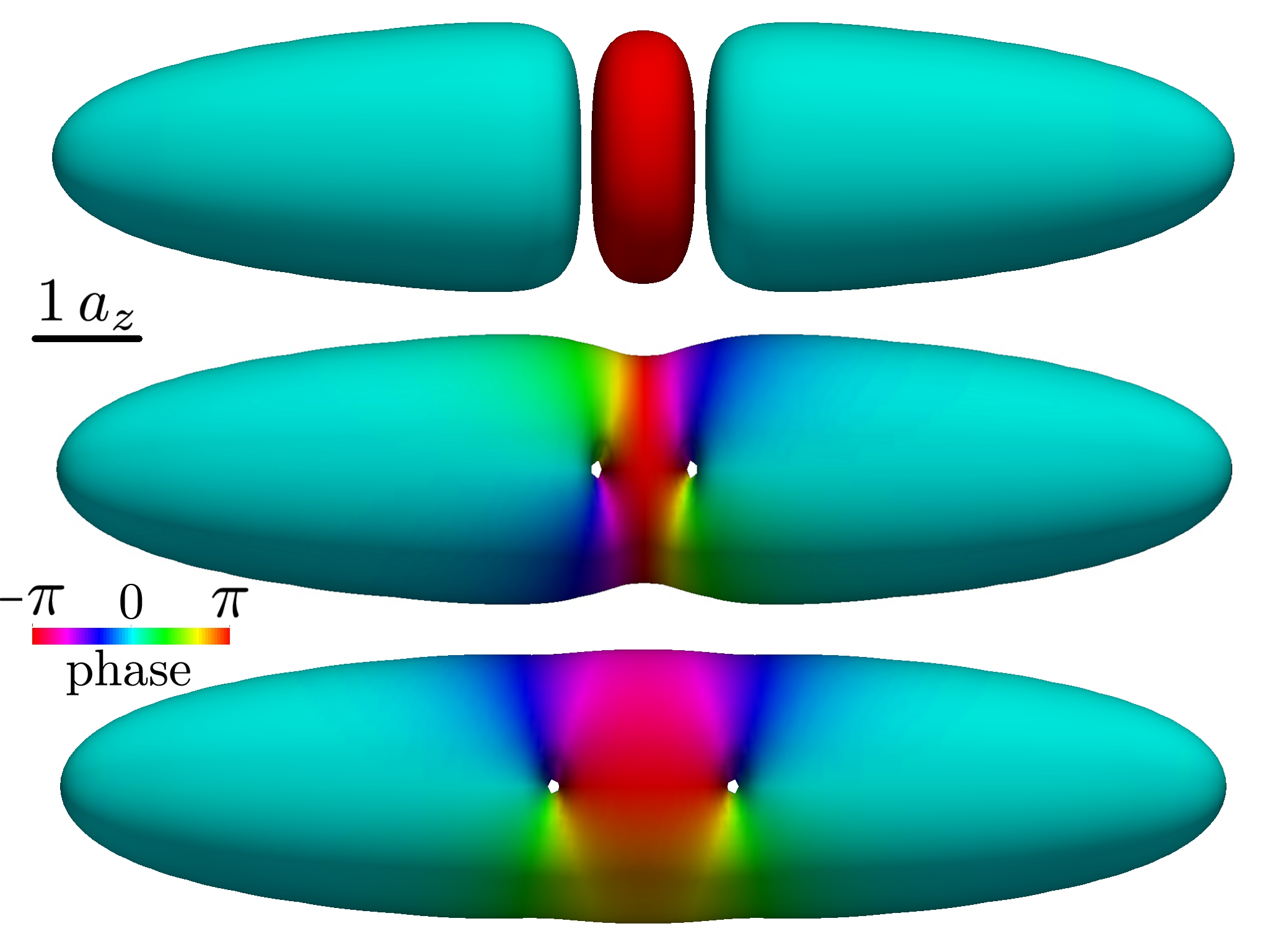}\\
\includegraphics[width=\linewidth]{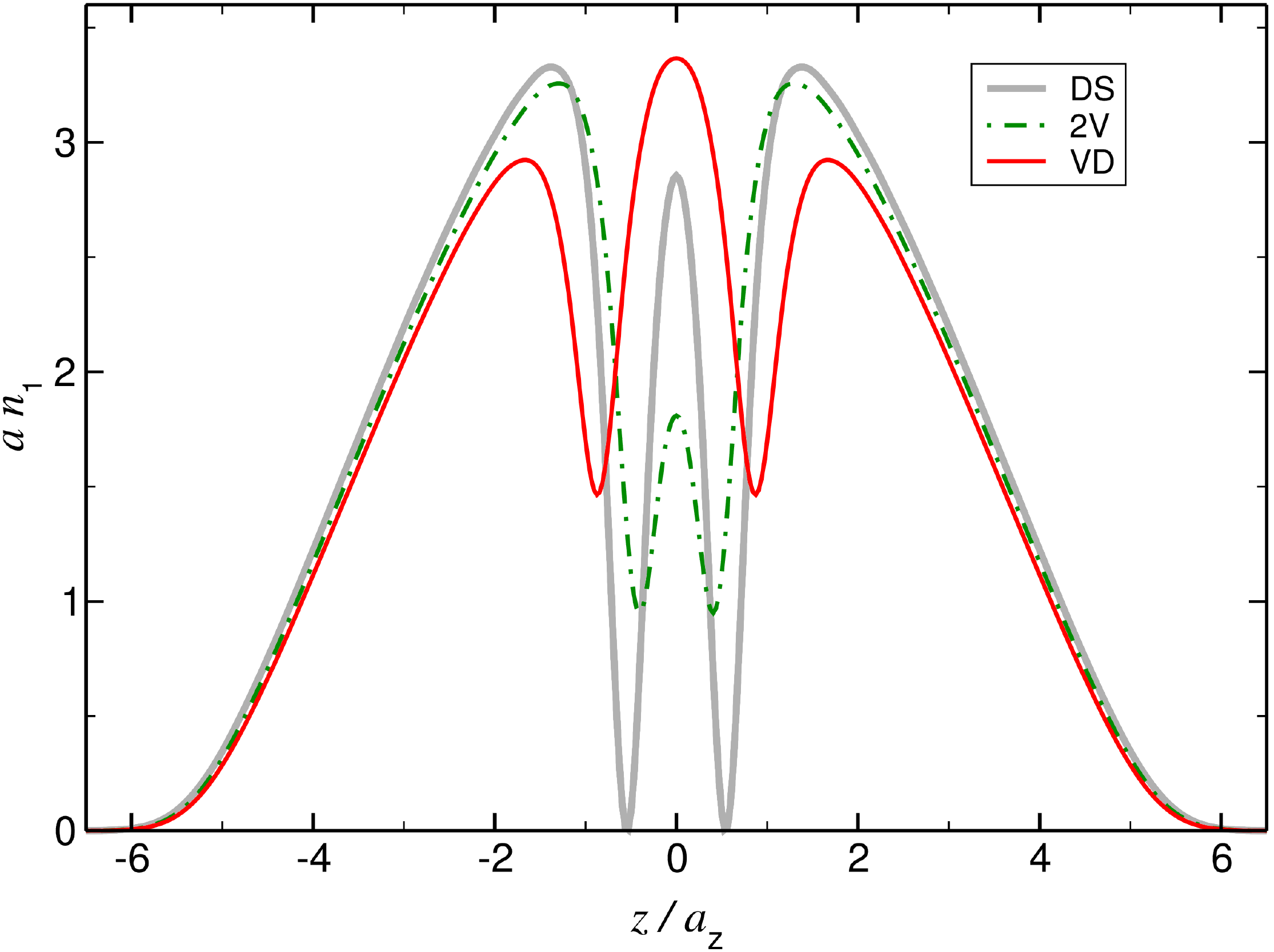}
\caption{Density isocontours (at 5$\%$ of the maximum density and colored by 
phase) of bound solitary waves in a harmonic trap with aspect 
ration $\lambda=5$. From top to bottom: two dark solitons (DS), two co-rotating 
vortices (2V) and two counter-rotating vortices (or vortex dipole VD). 
The three states are stationary solutions of the full GP 
equation for fixed interaction parameter $\chi=8.94$. The bottom panel depicts 
the non-dimensional axial density $a \,n_1(z)$ (see text) of these states. }
\label{Fig:3Dshape_e14}
\end{figure}
Once the snaking instability starts to operate in a state with high enough 
chemical potential, the solitons decay into vortex lines. The vortices are 
generated from the excitation of the unstable transverse modes available at 
that value of $\mu$, as depicted in Fig. \ref{Fig:3Dimag}.
Apparently, this is the same scenario found for single dark solitons 
\cite{MunozMateo2014}. However, the interaction 
between the emerging solitary waves introduces additional features that lead to 
a different outcome. First, not all the unstable frequencies of a 
bound-two-soliton state are pure imaginary, which causes the decay into 
non-static, oscillatory states \cite{Strogatz2018}. Some of the unstable modes 
with $l=0$ (dashed lines) depicted in Fig. \ref{Fig:3Dimag} belong to this set. 
Second, for given $l$ the 
instabilities appear in pairs, corresponding to the even and odd axial parity 
of the associated modes, with the even parity modes emerging at lower 
chemical potential than the odd ones (that present an axial node).
The prototypical example is the instability with 
$l=1$, which gives rise to bifurcations of a solitonic vortex in each soliton 
plane (see Fig. \ref{Fig:bifurcation}). In an isolated dark soliton all the 
transverse directions with 
the two possible circulations of the solitonic vortex are degenerate. In 
two bound solitons the orientations of the two emerging vortices are 
parallel, and the axial parity breaks the degeneracy between the 
counter-rotating configuration (VD, with even parity and lower energy) and the 
co-rotating one (2V, odd parity). As can be seen in  Fig. \ref{Fig:3Dimag}, the 
energy differences induced by the axial parity are reduced for larger aspect 
ratios.

\begin{figure}[tb]
\includegraphics[width=\linewidth]{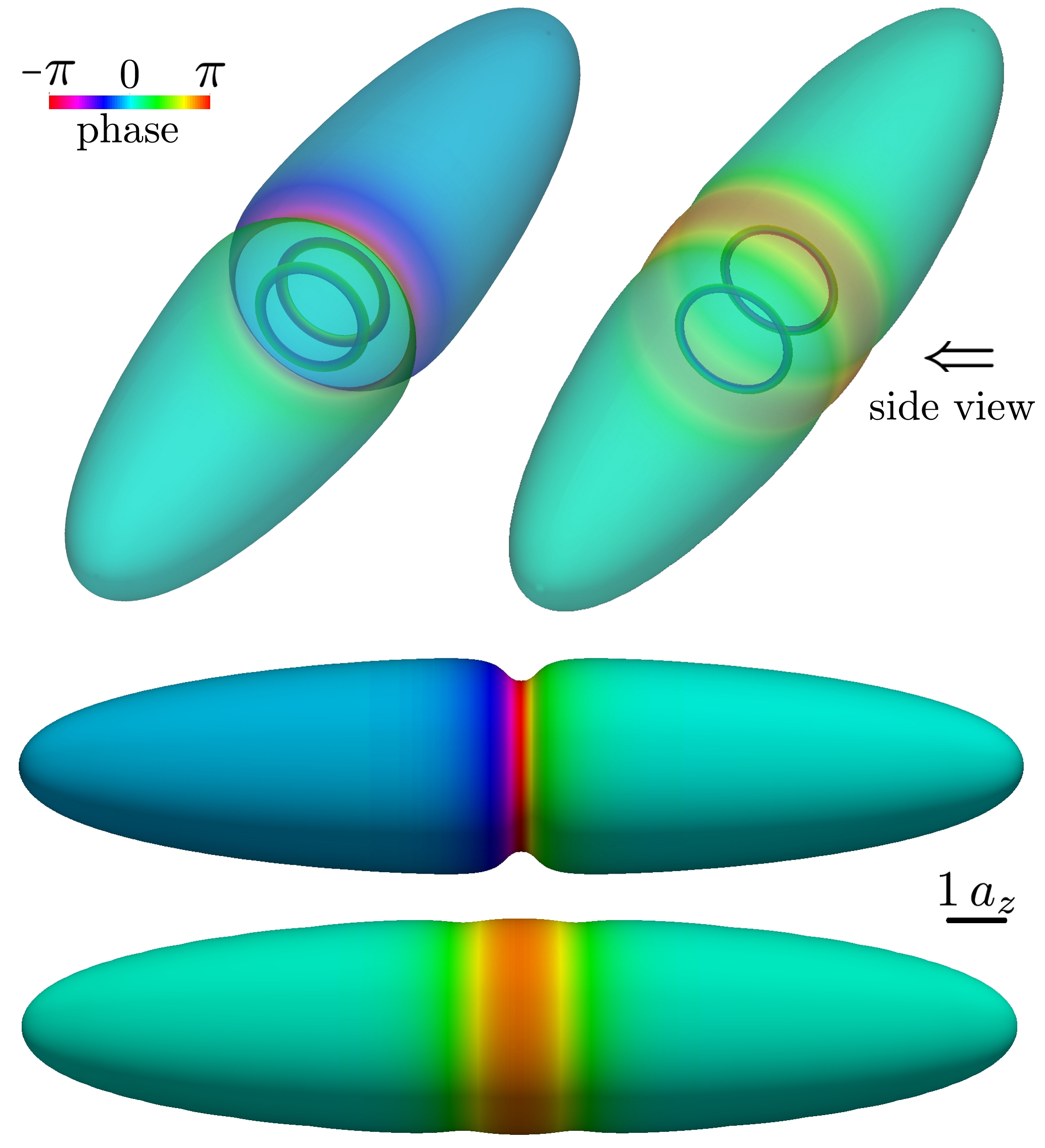}\\
\includegraphics[width=\linewidth]{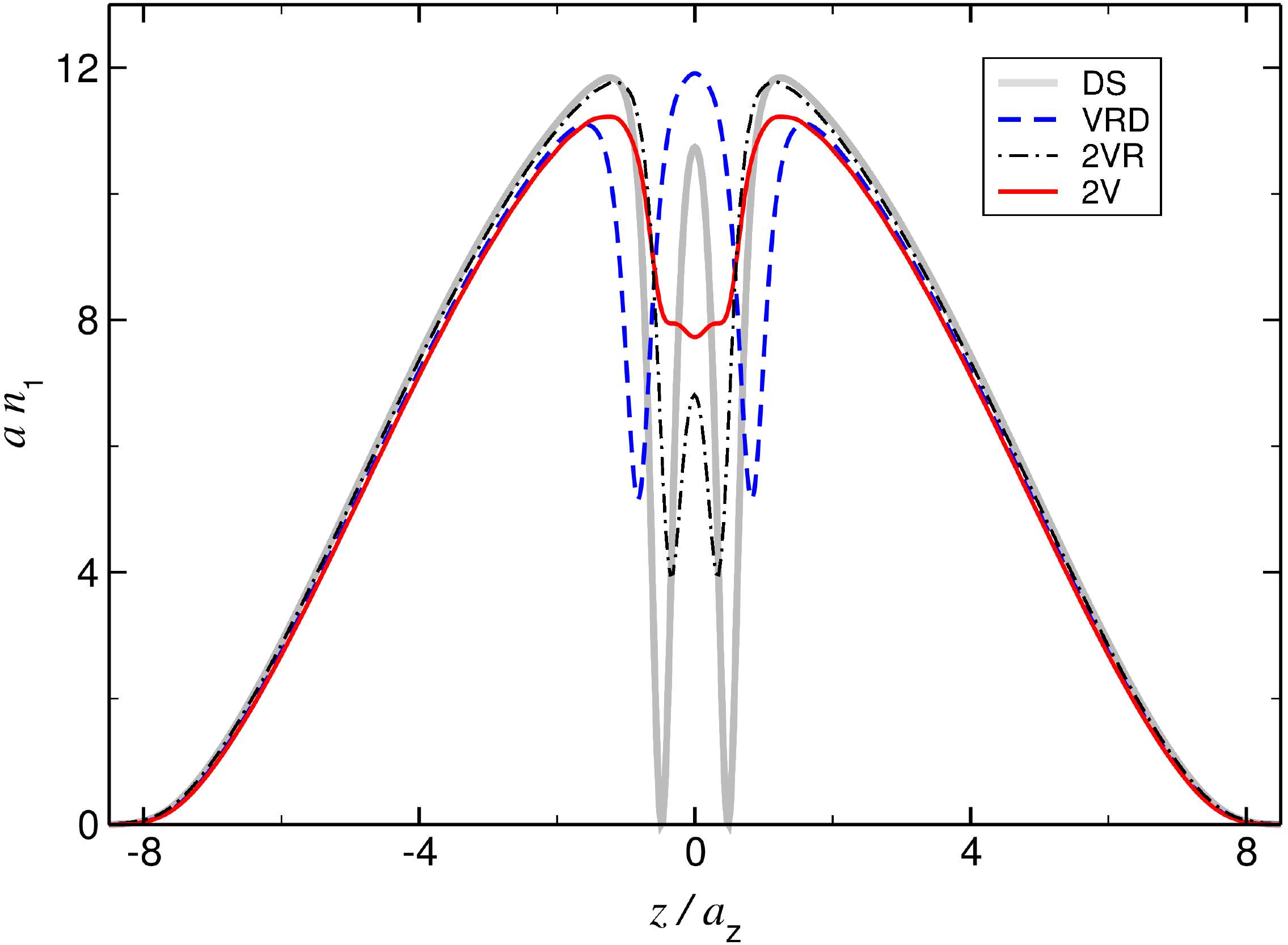}
\caption{Same as Fig. \ref{Fig:3Dshape_e14} for bound-vortex-ring 
states. The upper pictures (from left to right) show a 
semi-transparent perspective view, and the lower pictures (from top to 
bottom) show a side view of co-rotating vortex rings (2VR) and counter-rotating 
vortex rings (VRD), respectively, both at interaction $\chi=44.7$. The bottom 
panel, showing the axial densities $a \,n_1(z)$, includes the dark-soliton 
and co-rotating-vortex profiles (at equal interaction) for comparison.}
\label{Fig:3Dshape_e22}
\end{figure}
Figure \ref{Fig:bifurcation} shows the bifurcations of the first bound solitary 
waves from bound two solitons in a system with $\lambda=5$. The chemical 
potential of the solitonic states is shown versus interaction, which is 
parametrized by $\chi=N a/a_\perp\sqrt{\lambda}$. The bifurcations coincide 
with the emergence of unstable modes of same azimuthal polarity $l$ that 
possess pure imaginary frequencies (as shown in Fig. \ref{Fig:3Dimag}), whereas 
the complex frequencies lead to oscillatory dynamics.
Apart from the bound open-vortex families (VD and 2V), Fig. 
\ref{Fig:bifurcation} includes the bifurcation of bound counter-rotating 
vortex rings in a dipole configuration (VRD), where for each element of 
vortex line in one of the rings there is another element in the parallel ring 
which is its mirror image.

Representative examples of bound solitary waves are shown in Figs. 
\ref{Fig:3Dshape_e14} and ~\ref{Fig:3Dshape_e22}, obtained
 from the numerical solution of the 3D GP equation with interaction 
parameters $\chi=8.94$ and $44.7$, respectively.
The top panels represent the density isocontours (colored by phase) of the BEC 
at $5 \%$ of its maximum density, and the bottom panels depict the 
(non-dimensional) axial density profiles $an_1(z)= a\int 
dx\,dy\,|\psi(x,y,z)|^2$, after integration along the transverse coordinates.
Similar features to those discussed for 2D systems can be observed in the 
two-straight-vortex states of Fig. \ref{Fig:3Dshape_e14}, and in the 
two-vortex-rings shown in Fig.~\ref{Fig:3Dshape_e22},
where the inter-vortex separation is clearly larger for the vortex dipole 
configurations (VD and VRD). 
\begin{figure}[tb]
\includegraphics[width=\linewidth]{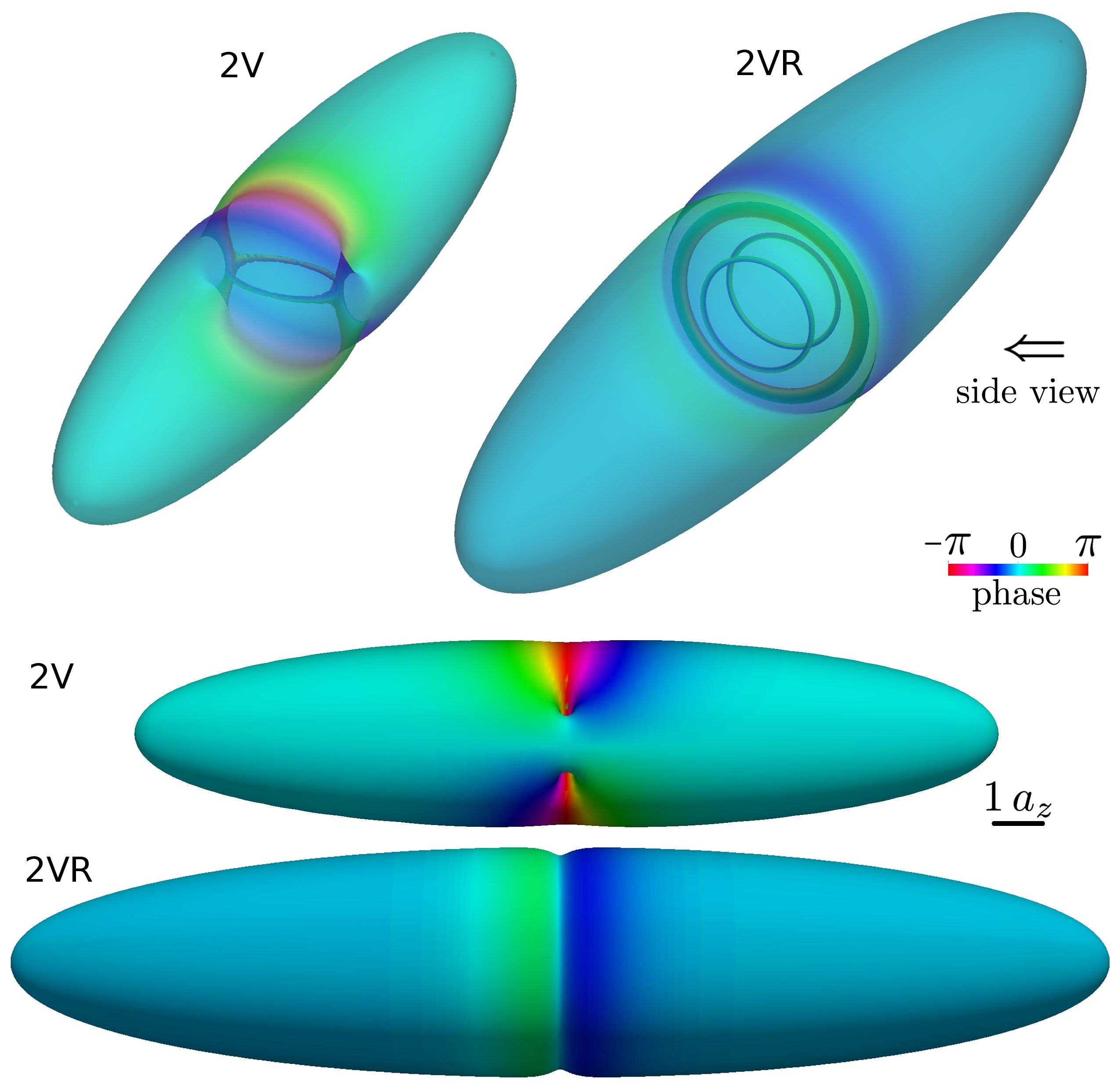}
\caption{Same as Fig. \ref{Fig:3Dshape_e22} for states of co-rotating vortex 
lines at higher interaction. The open-vortex configuration (2V) correspond to 
$\mu=4\,\hbar\omega_\perp$ (or $\chi=44.7$, same as states in Fig. 
\ref{Fig:3Dshape_e22}) and the vortex rings (2VR) to 
$\mu=11.4\,\hbar\omega_\perp$ (or $\chi=156.5$).}
\label{Fig:3D_complex}
\end{figure}

\begin{figure}[t]
\centering
\includegraphics[width=0.85\linewidth]{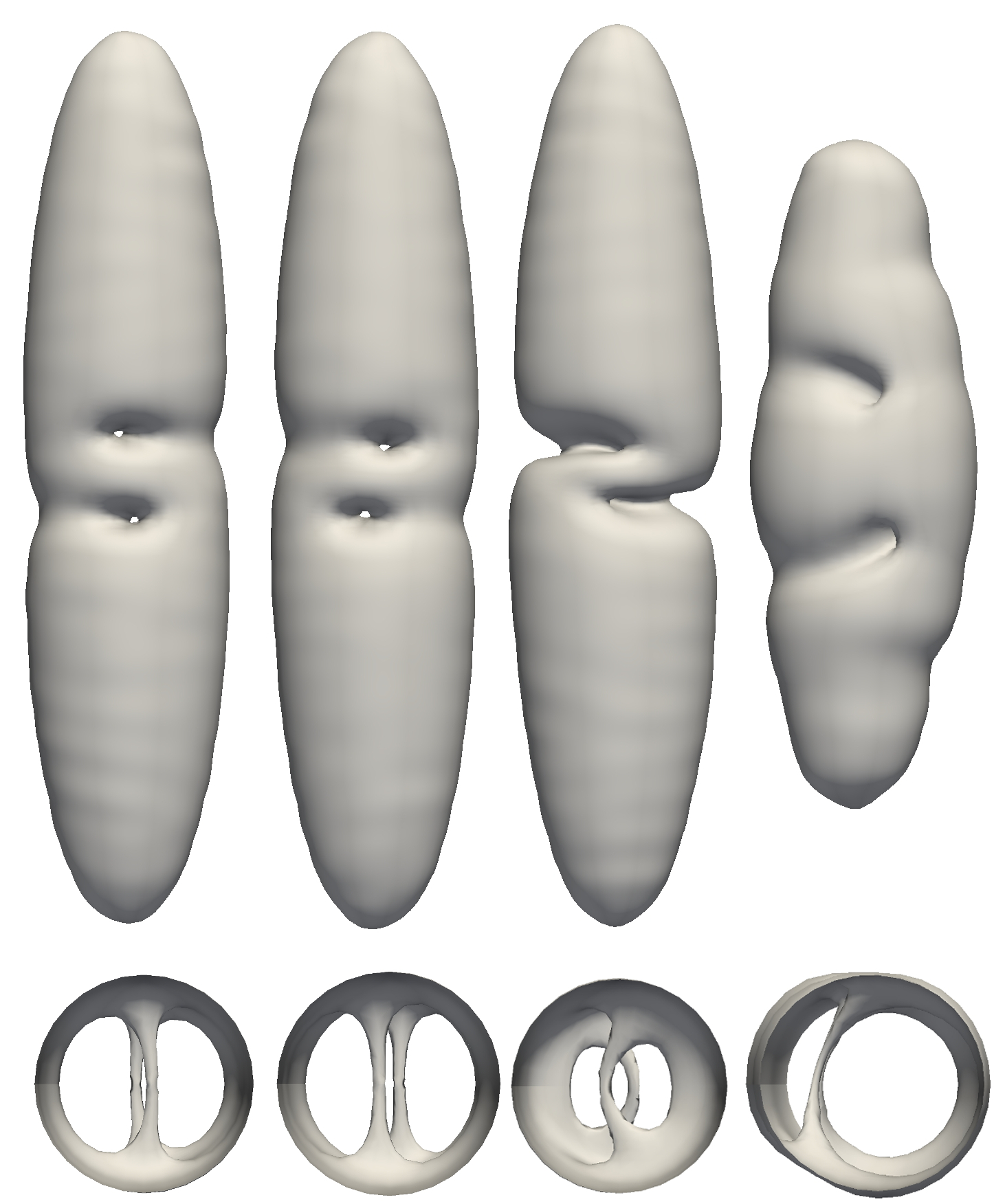}
\caption{Real time evolution of the two co-rotating-vortex 
state depicted in Fig. \ref{Fig:3Dshape_e14}, after the addition of 
perturbative white noise. The side views (top row) and the inner axial views 
(bottom row) of the density isocontours (at 5$\%$ of maximum density) are shown, 
from left to right, at times $t=19,\, 21.5, \, 23.5$ and $26$ ms.}
\label{Fig:2Vdecay}
\end{figure}
Further comments about the configurations of the states containing 
co-rotating vortices are in order. These states evolve, for increasing 
chemical potential, into complex 3D structures. Two examples are shown in 
Fig.~\ref{Fig:3D_complex} for the families of co-rotating open vortices (2V) 
and co-rotating vortex rings (2VR). 
As it was the case in 2D systems, these solitonic families present a minimum 
of the inter-vortex distance when measured on the axial density profile. 
However, the 3D states evolve with the chemical potential by increasing the 
bending of the initially straight vortex lines, in such a way that at the 
center of the condensate (thus at higher density) the inter-vortex separation 
is greater than close to the condensate surface (at lower density). Eventually, 
near the chemical potential value where the axial inter-vortex 
distance reaches the minimum, the end points of the two co-rotating vortices 
merge with two new perpendicular vortices (one at each end point of the initial 
vortices). This 3D configuration (visible in the perspective view (2V) of Fig. 
\ref{Fig:3D_complex}) reduces the otherwise increasing angular momentum for
increasing inter-atomic interaction. From this point on, the 
axial density profile (represented with a solid red line in the bottom panel of 
Fig. \ref{Fig:3Dshape_e22}) can not 
capture the features of the 3D structure, and a single thick density notch 
opens in the middle of the condensate. 

In a similar way, the family of two co-rotating vortex rings incorporate new 
vortex lines (a vortex ring larger and with opposite circulation) at the trap 
center. In this case, the new configuration reduces the axial linear 
momentum generated by the approaching off center rings. The resulting 
vortex pattern (visible in the perspective view (2VR) of Fig. 
\ref{Fig:3D_complex}) shwows a three vortex-antivortex-vortex sequence of rings.

A relevant difference with respect to the case of single solitary 
waves concerns the stability and dynamics of vortex states. We have only 
found stable states in the family of counter-rotating-straight vortices (VD). 
Stability is expected just after the bifurcation, inheriting 
this property from the parent bound-soliton 
states \cite{Stockhofe2011}. However, the bound-vortex states also inherit an 
oscillatory instability at higher chemical potential. As a result, for instance 
at $\lambda=5$, a state with $\mu=2.45\hbar\omega_\perp$ is stable while a state
with $\mu=3.25\hbar\omega_\perp$ is not. After this instability 
region, stability is recovered at even higher chemical potential. The VD
state with $\mu=4\hbar\omega_\perp$ shown in Fig. \ref{Fig:3Dshape_e14} is a 
stable example. In general, the 
typical decay of bound states show a 
complex vortex dynamics of rebounds or reconnections
\cite{Becker2013,Serafini2017}. We show an example of bound 
vortex decay in Fig. \ref{Fig:2Vdecay}, where selected snapshots during the 
real time evolution illustrate the process. The initial stationary state 
corresponds to the two-co-rotating-vortex configuration shown in Fig. 
\ref{Fig:3Dshape_e14}, upon which a peturbative white noise has 
been added. As can be seen, a small variations in the vortex positions are 
manifested at $t\approx 20$ ms. This initial departure from equilibrium 
leads eventually to the breakdown of the stationary flow created by the 
vortices, that show strong bending (visible in the inner views at the bottom 
of Fig. \ref{Fig:2Vdecay}). As a result, the particle density distribution 
becomes strongly distorted and the two solitonic vortices move apart.

\section{Conclusions}
\label{sec:conclusions}

In this work, we have analyzed static configurations of bound states made of 
solitary waves in harmonically trapped BECs. Motivated by 
a common setting in current experiments, we have focused 
on elongated, multidimensional condensates with realistic parameters. Therein 
solitonic states bifurcating from two bound dark solitons have been considered 
in 2D and 3D settings. We report on states containing either 
two-co-rotating or two-counter-rotating vortex lines 
having both a straight and a ring configuration. Among the families of 
solitonic states, only bound dark solitons and bound vortex 
dipoles have been found to support dynamically 
stable states, thus feasible to experimental realization. In the unstable 
regime of two bound dark solitons, the 
emergence of unstable modes with pure imaginary excitation frequencies marks 
the bifurcation of the static vortex states. Contrary to the case of a single 
dark soliton, we have also found (genuine 3D) unstable modes with non-vanishing 
real frequencies that give rise to oscillatory dynamics.

The analyzed states shed light on the nature of the 
soliton-soliton  and (solitonic) vortex-vortex interactions. Contrary to the 
untrapped systems, the inhomogeneous density profile 
induces repulsive inter-vortex interactions irrespective of the vortex 
circulations. In all cases, the associated repulsive forces balance the 
buoyancy forces  dragging the solitons and vortices towards the trap center. 
In the range of chemical potential analyzed, counter 
rotating vortices show  equilibrium distances that increase with the 
inter-atomic interaction. For co-rotating vortices this quantity shows a 
non-monotonic behavior that reflects a change in the vortex 
configuration: a 3D vortex-chain is shaped in the family of open vortex lines 
in order to arrest the increasing angular momentum, whereas a new 
counter-rotating vortex ring is introduced at the trap center in the family of 
vortex rings to reduce the axial linear momentum.

Natural extensions of the present work are envisaged. The evolution of 
the bound vortex families at higher chemical potential, or the search for 
additional analytical expressions for the equilibrium distances of solitons and 
vortices are issues that deserve further exploration.

\acknowledgments

M. G. and R. M. 
acknowledge financial support from Ministerio de Econom{\'i}a y 
Competitividad (Spain), Agencia Estatal de Investigaci\'on (AEI) and Fondo Europeo de Desarrollo Regional (FEDER, EU) 
under Grants No. FIS2014-52285-C2-1-P and FIS2017-87801-P, and from Generalitat de Catalunya Grant 
No.  2017SGR533.

\bibliography{bound_two_solitons}

\end{document}